\documentclass[useAMS,usenatbib]{mn2e}

\usepackage{graphicx}
\usepackage{natbib}
\newcommand{\HI}{H{\,\small I}}
\newcommand{\ltsima} {$\; \buildrel < \over \sim \;$}
\newcommand{\gtsima} {$\; \buildrel > \over \sim \;$}
\newcommand{\lta} {\lower.5ex\hbox{\ltsima}}
\newcommand{\gta} {\lower.5ex\hbox{\gtsima}}
\newcommand{\kmsMp}{km s$^{-1}$ Mpc$^{-1}$}
\newcommand{\kms}{km\ s$^{-1}$}
\newcommand{\FRI}{FR{-\small I}}
\newcommand{\FRII}{FR{-\small II}}
\newcommand{\OIII}{O{\,\small III}}
\newcommand{\HII}{H{\,\small II}}
\newcommand{\OI}{O{\,\small I}}
\newcommand{\OII}{O{\,\small II}}
\newcommand{\SII}{S{\,\small II}}
\newcommand{\NII}{N{\,\small II}}

\newcommand{\Cak}{Ca{\,\small II~}K}
\newcommand{\Cah}{Ca{\,\small II~}H}
\newcommand{\Ha}{H$\alpha$}
\newcommand{\Hb}{H$\beta$}

\title[Disc-dominated host galaxy of radio source B2~0722+30]{The disc-dominated host galaxy of FR-I radio source B2~0722+30}
\author[B. H. C. Emonts et al.]{B. H. C. Emonts$^{1,2}$\thanks{E-mail:bjorn.emonts@csiro.au}, C. N. Tadhunter$^{3}$, R. Morganti$^{4,5}$, T. A. Oosterloo$^{4,5}$,
\newauthor J. Holt$^{6}$, E. Brogt$^{7}$, G. van Moorsel$^{8}$.\\
$^{1}$CSIRO Australia Telescope National Facility, PO Box 76, Epping NSW, 1710, Australia\\
$^{2}$Department of Astronomy, Columbia University,  Mail Code 5246, 550 West 120th Street, New York, N.Y. 10027, USA\\ 
$^{3}$Department of Physics and Astronomy, University of Sheffield, Sheffield S3 7RH, UK\\
$^{4}$Netherlands Foundation for Research in Astronomy, Postbus 2, 7990 AA Dwingeloo, the Netherlands\\
$^{5}$Kapteyn Astronomical Institute, University of Groningen, P.O. Box 800, 9700 AV Groningen, the Netherlands\\
$^{6}$Leiden Observatory, Leiden University, P.O. Box 9513, 2300 RA Leiden, the Netherlands\\
$^{7}$The University of Arizona, Steward Observatory, 933 North Cherry Avenue, Tucson, AZ 85721, USA\\ 
$^{8}$National Radio Astronomy Observatory, Socorro, NM 87801, USA\\
}
\begin{document}

\date{}

\pagerange{\pageref{firstpage}--\pageref{lastpage}} \pubyear{2007}

\maketitle

\label{firstpage}

\begin{abstract}
We present new observational results that conclude that the nearby radio galaxy B2~0722+30 is one of the very few known disc galaxies in the low-redshift Universe that host a classical double-lobed radio source. In this paper we use \HI\ observations, deep optical imaging, stellar population synthesis modelling and emission-line diagnostics to study the host galaxy, classify the Active Galactic Nucleus and investigate environmental properties under which a radio-loud AGN can occur in this system. Typical for spiral galaxies, B2~0722+30 has a regularly rotating gaseous disc throughout which star formation occurs. Dust heating by the ongoing star formation is likely responsible for the high infra-red luminosity of the system. The optical emission-line properties of the central region identify a Low Ionization Nuclear Emission-line Region (LINER)-type nucleus with a relatively low [\OIII] luminosity, in particular when compared with the total power of the Fanaroff $\&$ Riley type-I radio source that is present in this system. This classifies B2~0722+30 as a classical radio galaxy rather than a typical Seyfert galaxy. The environment of B2~0722+30 is extremely \HI-rich, with several nearby interacting galaxies. We argue that a gas-rich interaction involving B2~0722+30 is a likely cause for the triggering of the radio-AGN and/or the fact that the radio source managed to escape the optical boundaries of the host galaxy.

\end{abstract}


\section{Introduction}
\label{sec:introduction}

Radio galaxies are systems that contain an Active Galactic Nucleus (AGN) from which powerful radio-continuum jets emerge. In many cases, these radio jets reach far beyond the optical boundaries of the host galaxy. In the nearby Universe, the host galaxies of classical radio sources are generally early-type systems. \citet{ver01} argue that almost all classical low-$z$ double radio sources are associated with giant elliptical galaxies and that the more discy S0 hosts are often misidentified ellipticals. In the hierarchical model of galaxy formation, early-type galaxies form the end products of merging systems. Indeed, \citet{hec86} and \citet{smi89} showed that a significant fraction of the more powerful, edge-brightened radio galaxies of Fanaroff $\&$ Riley type-II \citep[\FRII;][]{fan74} still show optical signatures of the merger event (such as tails, bridges, shells, etc.). Lower power, edge-darkened Fanaroff $\&$ Riley type-I (\FRI) radio sources are generally also hosted by genuine elliptical galaxies, but signs of a merger event in these systems are often much more subtle \citep[e.g.][]{col95} or not apparent at all \citep[e.g.][]{hec86,bau92,emo07}. Recent studies suggest that the difference in host galaxy properties between various types of radio sources may be more closely related to the excitation properties of the AGN rather than merely the FR-type \citep{all06,har07,bal08}, but the fact remains that radio galaxies in the nearby Universe are typically early-type systems. Also for low redshift QSOs, \citet{smi86} find a trend that the host galaxies of radio-loud objects tend to be better fitted by elliptical galaxy models and show indications that they are more likely to have optical peculiarities than radio-quiet hosts, which are better fitted by disc galaxy models. However, in a more recent study \citet{dun03} conclude that, for nuclear luminosities above the traditional quasar-Seyfert boundary of $M_{\rm V}\rm{(nuc)} \sim -23$, all low-redshift quasars (both radio-loud and radio-quiet) are hosted by massive ellipticals.

In contrast, AGN in spiral galaxies are typically Seyfert nuclei \citep{sey43}, which contain a bright emission-line region as a result of an active nucleus that is luminous in the optical, ultra-violet and X-ray (though not as luminous as typical QSOs/quasars). Seyfert galaxies often also contain a nuclear radio source, but in contrast to radio galaxies these sources are generally weak and very compact \citep[e.g.][and references therein]{mid04}. Radio-loud Seyferts (i.e. with radio powers in the regime of radio galaxies) do exist, but they are rare and their radio component generally does not reach beyond the nuclear region of the host galaxy \citep[e.g.][and references therein]{kom06}.

Although it is very rare for genuine disc galaxies in the nearby Universe to host an extended radio source, two clear counter-examples exist that have been studied in detail; the spiral galaxy 0313-192 in the Abell cluster A428 hosts a giant (350 kpc) double-lobed \FRI\ radio source \citep{led98,led01,kee06}, while NGC~612 is a typical S0 galaxy with a large-scale star-forming \HI\ disc and a powerful \FRI/\FRII\ radio source \citep[][see Sect. \ref{sec:wrong} for more details on these two systems]{ver01,emo08}. A third example of a spiral-hosted radio galaxy is the BL Lac object PKS~1413+135 \citep{mch94,lam99}, but the properties of its radio source are difficult to investigate, given that the radio jets are pointing roughly in the line-of-sight of the observer. These rare cases of disc-dominated radio galaxies provide an excellent opportunity for studying host galaxy properties and environmental effects that could be important for the triggering and/or evolution of the radio source in these systems. This could provide a better insight in the origin and evolution of radio-loud AGN in general. In addition, a detailed knowledge of these nearby disc-dominated radio galaxies provides valuable information for comparison studies at high redshift, for which recent results suggest that spiral host galaxies may be much more common among classical radio sources than at low $z$ \citep{nor08}.

In this paper we study another nearby disc-dominated galaxy that hosts an extended radio source, namely B2~0722+30. In the literature, B2~0722+30 has occasionally been mentioned as a disc-dominated radio galaxy, but it has not been discussed in detail, mainly because its AGN is often confused with that of Seyfert galaxies. In this paper we investigate the optical and radio properties of the AGN in B2~0722+30 in order to obtain a more reliable classification. We also study the morphology, gas content and star formation properties of the host galaxy and its environment. This allows us to verify whether B2~0722+30 is indeed a rare example of a disc-dominated radio galaxy and investigate the conditions under which an extended radio source can occur in this system.

\subsection{B2~0722+30}

The radio source B2~0722+30 has a two-sided jet with a slightly distorted structure \citep{rui84,fan86}. Being associated to a nearby galaxy ($z = 0.0189$), its total radio power is $P_{\rm 408 MHz} = 1.3 \times 10^{24}$  W Hz$^{-1}$ \citep{gio05}\footnote{corrected for $H_{0} = 71$ \kmsMp.}. Although this is at the low end of the radio continuum power for radio-loud AGN, the radio structure is 13.6 kpc in extent, reaches outside the optical host galaxy and has an \FRI\ morphology \citep{par86,fan86,fan87}. From {\em Hubble Space Telescope (HST)} imaging, \citet{cap00} claim that B2~0722+30 is hosted by a spiral galaxy. Dust lanes are clearly visible along the disc and bulge of this system. A CO detection tracing $1.2 \times 10^{9} M_{\odot}$ of H$_{2}$ gas in B2~0722+30 was discovered by \citet{maz93} and followed-up by \citet{eva05} using single-dish observations. In a recent paper \citep{emo07}, we presented a preliminary analysis of the neutral hydrogen (\HI) gas in B2~0722+30, which shows that an \HI\ disc follows the optical disc in this system. A more detailed \HI\ analysis is given in this paper. 

The outline of the paper is as follows. In Sect. \ref{sec:observations} we describe the new observations presented in this paper. Section \ref{sec:results} investigates the properties of the AGN, the host galaxy and the environment of B2~0722+30 (optical morphology, gas content, star formation properties and AGN line diagnostics). This is followed in Sect. \ref{sec:discussion} by a detailed discussion about the classification of B2~0722+30 as a disc-dominated radio galaxy and the conditions under which an extended radio source occurs in this system. In Sect. \ref{sec:conclusions} we give the final conclusions.\\
\vspace{-2mm}\\
For simplicity, in this paper, we will use the name B2~0722+30 for both the radio source and the host galaxy. We use $H_{0} = 71$ \kmsMp\ throughout this paper, which puts B2~0722+30 at a distance of 80 Mpc (1 arcsec = 0.39 kpc). 

\section{Observations}
\label{sec:observations}

\subsection{Neutral hydrogen}
\label{sec:obshi}

\HI\ observations of B2~0722+30 were done with the Very Large Array in C-configuration on 23 December 2002. The effective integration time on the source was 3.1 h. We used the 6.25 MHz band with 64 channels in 2 IF mode. For the data reduction and visualisation we used MIRIAD and KARMA. After flagging and calibration, we fitted a straight line to the line-free channels in order to separate the continuum from the line data. Even at uniform weighting the radio source in B2~0722+30 is only marginally resolved. For a detailed image of the radio continuum of B2~0722+30 we refer to \citet[][also shown in Fig. \ref{fig:jetalignment}]{fan86}. The line data were used to construct a natural weighted data-cube with a beam of $19.0 \times 17.5$ arcsec$^{2}$, velocity resolution of 43 \kms\ (after Hanning smoothing) and a noise level of 0.19 mJy beam$^{-1}$. From this data set a mask was created by smoothing the data spatially by a factor 1.4 and subsequently masking out all the signal below 3$\sigma$ plus discarding the signal from the absorption. This mask was used to extract a total intensity image of the emission-line gas in the data cube by adding the signal in the regions that were not masked out (Fig. \ref{fig:HI_B20722}). In order to visualise in detail the signal in the separate channels (Fig. \ref{fig:chanmaps0722}), another data-cube was made with robust +1 weighting and smoothed to a beam-size of $26.5 \times 24.4$ arcsec$^{2}$. The velocity resolution of this data cube is 43 \kms\ (after Hanning smoothing) and the noise level is 0.23 mJy beam$^{-1}$.

\subsection{Optical imaging}
\label{sec:obsoptim}

A 3600 sec exposure B-band image of B2~0722+30 and its environment was obtained on 13 March 2007 at the 2.4m Hiltner telescope of the MDM observatory, located at the southwestern ridge of Kitt Peak, Arizona (USA). Observations were done under photometric conditions, at an airmass of sec $z < 1.2$ and with a seeing of about 1.1 arcsec. We used the Image Reduction and Analysis Facility (IRAF) to perform a standard data reduction (bias subtraction, flat-fielding, frame alignment and cosmic-ray removal). Likely due to minor shutter issues during readout, a gradient was present in the background of the $9.5 \times 9.5$ arcmin Echelle CCD image. We were able to remove this effect to a high degree by fitting the background in the region surrounding B2~0722+30 and subsequently subtracting this background-gradient from the image. Using KARMA, we applied a world coordinate system to the image by identifying a few dozen of the foreground stars in a Sloan Digital Sky Survey (SDSS) image of the same region. The newly applied coordinate system agrees with that of the SDSS image to within 1 arcsec, good enough for an accurate overlay with our \HI\ data.

\subsection{Optical spectroscopy}
\label{sec:obsoptspec}

Optical long-slit spectra of B2~0722+30 were taken at the William Herschel Telescope (WHT) on 12 January 2004 using the ISIS long-slit spectrograph with the 6100 \AA\ dichroic, the R300B and R316R gratings in the blue and red arm and the GG495 blocking filter in the red arm to cut out second order blue light. This resulted in a wavelength coverage from about 3500 to 8000 \AA\ and $\lambda$-resolution of about 5 \AA. The slit had a width of 1.3 arcsec and was aligned along the major axis of the host galaxy (PA 141$^{\circ}$). Observations were done at an airmass of about 1.0 and seeing conditions varied from 1.2 to 1.6 arcsec during the observations. The total integration time was 6000 sec for both the blue and red arm. We used IRAF for a standard reduction of the data (bias subtraction, flatfielding, wavelength calibration, cosmic-ray removal, background subtraction and tilt removal). Arc exposures for wavelength calibration were taken before and after the observations of B2~0722+30 at approximately the same position. We used a 60 sec observation of the telluric standard star HD58336 to correct for atmospheric absorption bands in the red part of the spectrum as well as possible. The resulting spectra are aligned within one pixel in the spatial direction. For the flux calibration we used five standard calibration stars (D191-B2B, Feige 67, PG0216+032, Feige 24 and HD19445). The accuracy of the relative flux calibration is within 6$\%$ (only in the very red part of the spectrum, beyond 7400 \AA, the flux calibration errors are up to 10$\%$). Due to slit-losses when observing the flux-calibration stars we could not normalise the spectrum better than within a factor of 2. We note that this does not affect the stellar population analysis of Sect. \ref{sec:stelpop} (because the relative flux calibration is unaffected), but it does introduce a factor of 2 uncertainty in the total stellar mass estimates derived from the spectra (Sect. \ref{sec:stellarmasses}). Subsequently, we used the Starlink package FIGARO to correct the spectra for galactic extinction (assuming $E(B-V)$ = 0.059)\footnote{Based on results from the NASA/IPAC Extragalactic Database (NED)} and to de-redshift the spectra to rest-wavelengths. For the analysis of the spectra we used the Starlink package DIPSO for emission-line fitting and the Interactive Data Language (IDL) for the stellar population modelling.

\subsubsection{Stellar population synthesis method}
\label{sec:SPmethod}

To investigate in detail the stellar populations in B2~0722+30, we model the continuum Spectral Energy Distribution (SED) of the optical spectra in various regions, taking into account both stellar and AGN-related components. Subsequently, we will make a more detailed comparison between the data and the models by investigating the age-sensitive \Cak\ and Balmer absorption lines in order to constrain our result better. This technique has been used before \citep[see][]{tad02,tad05,wil02,wil04} and is preferred to using absorption line indices at face-value, because most of the age sensitive diagnostic absorption lines could be affected by emission-line contamination due to the activity in B2~0722+30. 

For modelling the observed spectra, stellar population models from \citet{bru03} are used. These are based on Salpeter initial mass function (IMF) and solar metallicity, instantaneous starbursts. We use a $\chi^{2}$ minimisation technique to fit combinations of an unreddened 12.5 Gyr old stellar population (OSP), a young stellar population (YSP) and a power-law component of the form \(F_{\lambda} \propto \lambda^{\alpha}\) to the observed spectra (see Sect. \ref{sec:stelpop} for more details). YSP template spectra across different ages and with various amounts of reddening are used for this. The latter is necessary, since we were not able to determine the reddening a priori using \Ha/\Hb\ in B2~0722+30, because \Hb\ absorption appears to be significant in this system, which dilutes the \Hb\ emission line (and apart from \Ha, the other Balmer lines are too weak to be used). The \citet{sea79} reddening law is used to apply the reddening to the template spectra. The final library of YSP template spectra comprises YSP models with age 0.01 - 5.0 Gyr and with $E(B-V)$ = 0.0 - 1.6. We compare the total flux of the combined OSP plus YSP (plus possible power-law component) model with the observed flux in wavelength-bins across the spectrum. In order to scale the template models to the observed spectra, a normalising bin was chosen in the wavelength range 4720 - 4820 \AA. For the $\chi^{2}$-fitting we assume an error of $\pm$6$\%$ in each wavelength bin, in agreement with the uncertainty in the flux calibration. Note that, since the flux calibration errors are not likely to be independent between the various wavelength bins, we can merely use the reduced $\chi^{2}$ values as an indication of the region of parameter space for which the modelling provides good results, rather than derive accurate statistical properties of the fitting procedure itself. For that, we also need to inspect the model-fit to our spectra visually. The \Cak\ ($\lambda$ 3933 \AA) and higher order Balmer absorption lines in the blue part of the spectrum, as well as the G-band feature ($\lambda$ 4305 \AA), are excellent to use for this.

As investigated by \citet[][and references therein]{dic95}, spectra of active galaxies may be diluted by UV-excess due to nebular continuum and AGN related effects in the region shortward of 4000 \AA\ \cite[see also discussion in][]{tad02}. In a previous paper \citep{emo06} we investigated in detail the effects of nebular continuum in the spectral light of radio galaxy B2~0648+27, which was observed during the same observing run and under the same conditions as B2~0722+30. We concluded that for the nuclear spectrum of B2~0648+27 the most extreme cases - maximum nebular continuum subtraction and no nebular continuum subtraction - did not significantly alter our main results of the SED modelling. The emission-line equivalent width provides an indication for the contribution of nebular continuum \citep{dic95} as well as scattered AGN components \citep[see e.g.][]{tad02}. B2~0722+30 does not contain emission lines that are significantly stronger than the emission lines in B2~0648+27. We therefore do not expect that nebular continuum will have a significant contribution to the light in the spectra of B2~0722+30, hence no correction for the nebular continuum is applied.

\section{Results}
\label{sec:results}

\begin{figure*}
\centering
\includegraphics[width=\textwidth]{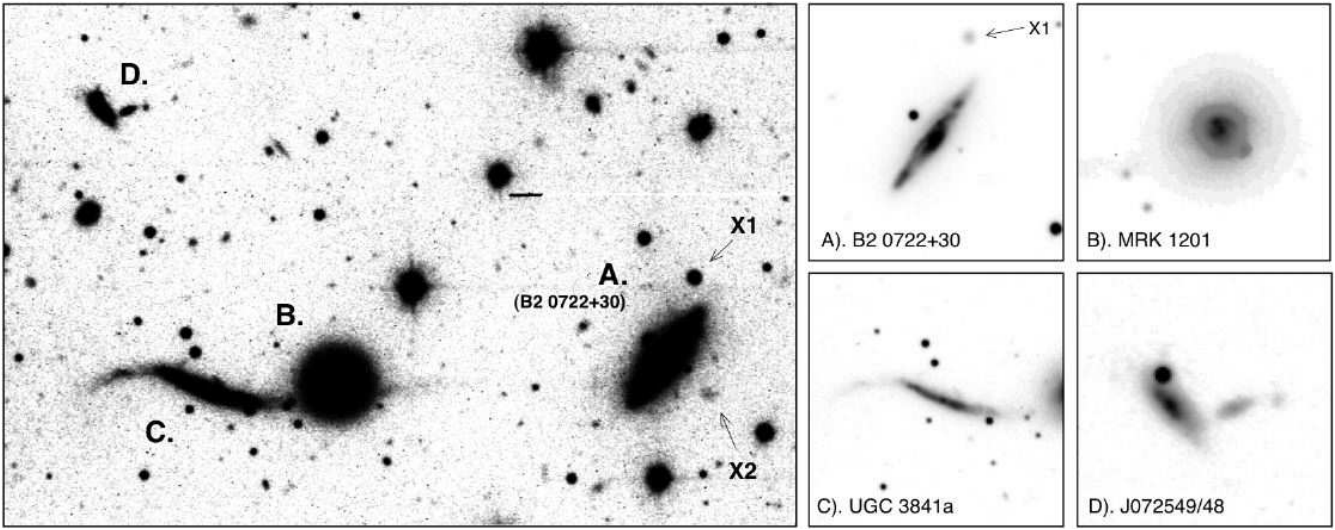}
\caption{Deep optical B-band image of B2~0722+30 and its environment. The high-contrast plot {\sl (left)} marks various galaxies and features of interest that are plotted in low contrast on the right and which are discussed in more detail in the text.}
\label{fig:Bband_B20722}
\end{figure*}

Figure \ref{fig:Bband_B20722} shows the deep B-band image of B2~0722+30 and its environment. B2~0722+30 (galaxy 'A') is clearly dominated by a stellar disc. The optical {\sl HST} imaging by \citet{cap00} shows in more detail the bulge component and dust features. Several other interesting galaxies and optical features are visible in the environment of B2~0722+30 in Fig. \ref{fig:Bband_B20722}. They have a striking connection with the neutral hydrogen gas in the region around B2~0722+30 and will therefore be discussed in detail in the next Section.

\subsection{Neutral and ionised gas}

\subsubsection{Host galaxy} 
\label{sec:host}

\begin{figure*}
\centering
\includegraphics[width=\textwidth]{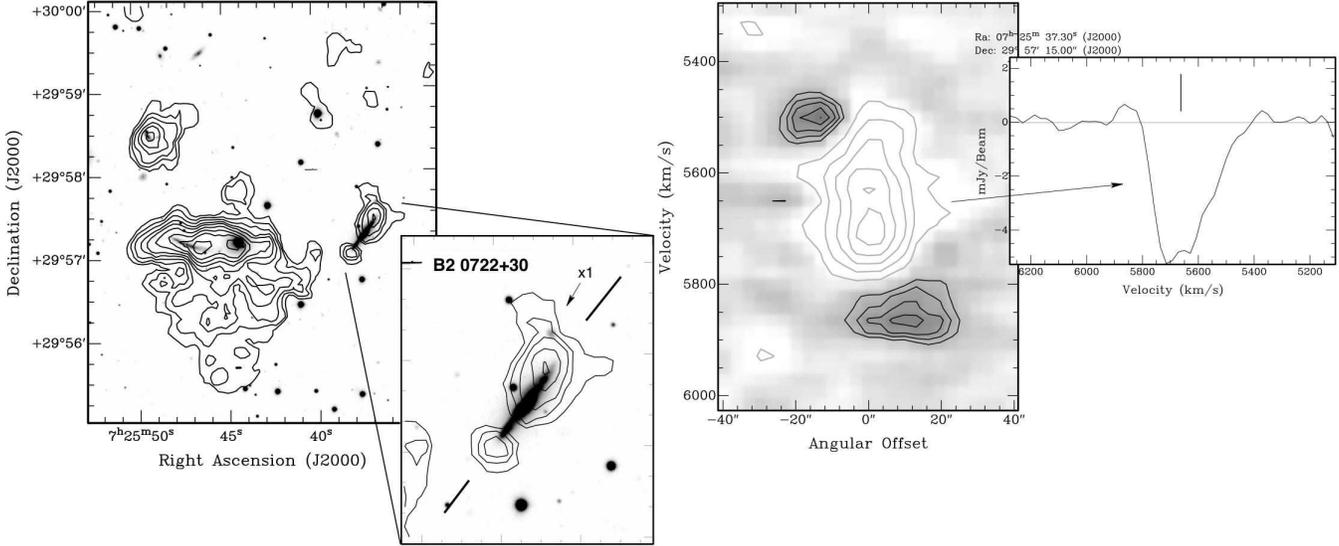}
\caption{{\sl Left:} total intensity map of \HI\ emission in B2 0722+30 and its environment (contours) overlaid on to the optical MDM image (grey-scale). Contour levels: 0.63, 1.2, 1.8, 2.7, 3.3, 4.3, 5.3, 6.3, 7.7 $\times 10^{20}$ cm$^{-2}$. The zoom-in shows B2 0722+30 in detail (for clarity, the part of the gas-disc that is seen in absorption against the radio source is not shown). {\sl Right:} Position-velocity plot along the major axis of the \HI\ disc, as indicated in the zoom-in on the left. Contour levels: -0.5 -1.4 -2.4 -3.4 -4.4 (grey); 0.5 0.7 0.9 1.1 1.3 (black) mJy beam$^{-1}$. The profile of the \HI\ absorption against the radio continuum is also show. The bar indicates the systemic velocity.}
\label{fig:HI_B20722}
\end{figure*}

\begin{figure*}
\centering
\includegraphics[width=\textwidth]{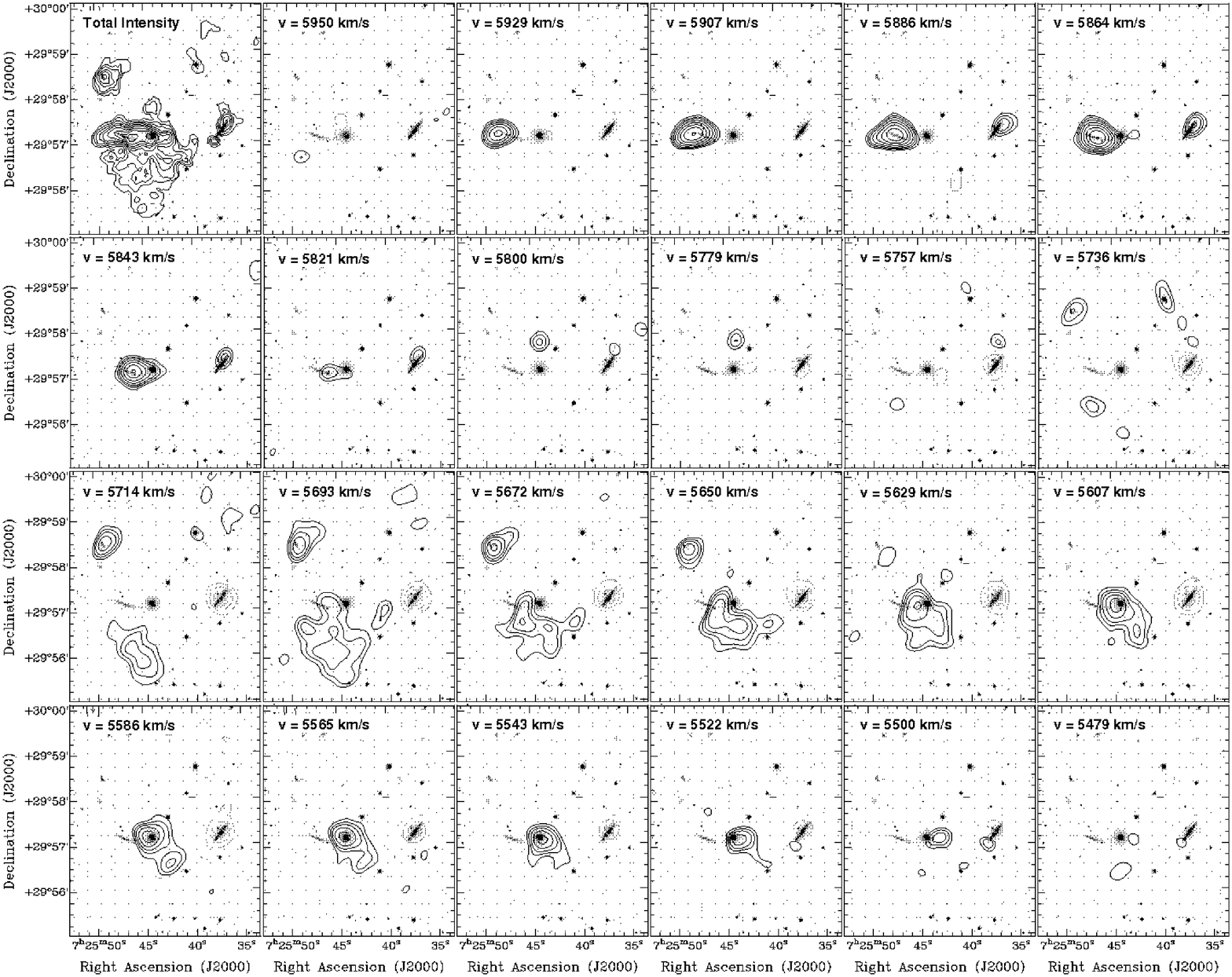}
\caption{Channel maps of the environment of B2~0722+30 from the smoothed data cube. The \HI\ bridge between MRK~1201 and B2~0722+30 is visible in the channels with $5714 \leq {\rm v} \leq 5650$ \kms. Contour levels: -0.7, -1.7, -3.0, -4.5 (grey); 0.7, 0.9, 1.1, 1.4, 1.8, 2.3, 2.9, 3.7 (black) mJy beam$^{-1}$. The first panel shows the total intensity image of Fig. \ref{fig:HI_B20722} (i.e. higher resolution than the channel maps).}
\label{fig:chanmaps0722}
\end{figure*}

Figure \ref{fig:HI_B20722} shows a total intensity map of the \HI\ gas detected in B2~0722+30 and its environment, while Fig. \ref{fig:chanmaps0722} shows the channel maps of the \HI\ data. The \HI\ gas in B2~0722+30 is distributed in a rotating disc that follows the optical disc of the host galaxy out to a radius of about 8 kpc from the centre. The \HI\ mass detected in emission in this disc is $M_{\rm HI} = 2.3 \times 10^{8} M_{\odot}$ and the average surface density is about 4.1 $M_{\odot}\ {\rm pc}^{-2}$. The total \HI\ mass in the disc will be somewhat larger than this, because the part of the \HI\ gas that lies in front of the radio continuum source is detected in absorption and not included in this \HI\ mass estimate. The \HI\ disc seems to be in regular rotation and covers a velocity range of 450 \kms, which is equal to the FWHM of the CO emission-line profile detected by \citet{eva05}. The \HI\ disc is centred on 5675 \kms. This is in agreement with the central part of the absorption as well as with redshift determinations from CO data by \citet{maz93} and \citet{eva05}.

Figure \ref{fig:0722apertures} shows a plot of the H$_{\alpha}$+[\NII]$\lambda$$\lambda$6548,6583 emission lines that we detect along the major axis of B2~0722+30. Also here, regular rotation is clearly visible. We trace the emission lines out to about 5.4 kpc on either side of the nucleus. The optical emission lines cover total  velocity range of $322 \pm 11$ \kms\ and are centred on a redshift velocity of $5666 \pm 60$ \kms (which corresponds to $z = 0.0189 \pm\ 0.0002$). We assume this is the systemic velocity of B2~0722+30, which is in good agreement with the central velocity of the \HI\ and CO gas described above.   

At the northwestern tip of the disc there appears to be a tail of \HI\ gas stretching to the north, in the direction of an optical counterpart (marked as ``X1'' in Figs. \ref{fig:Bband_B20722} and \ref{fig:HI_B20722}). The radial profile of this optical counterpart (deduced with the interactive graphics anaysis tools in IRAF) is significantly broader than that of the stars in the field. We therefore argue that this counterpart likely represents a small galaxy rather than a foreground star. If it is a galaxy, it can be either a true companion of B2~0722+30 or a galaxy in projection. Up to 48 kpc further out to the north (around dec=29$^{\circ}$59') there appears to be more faint \HI\ emission, although this has to be confirmed with additional data. Another, even fainter, peculiar optical feature is present about 6 kpc southwest of B2~0722+30. This feature is only visible in the deep optical image of Fig. \ref{fig:Bband_B20722} ("X2"). The nature of this very faint optical feature is not clear.

\subsubsection{Environment}
\label{sec:environment}

The environment of B2~0722+30 shows a number of luminous galaxies that appear to be in interaction. The most interesting is the galaxy pair UGC~3841 east of B2 0722+30. Our deep optical B-band image (Fig. \ref{fig:Bband_B20722}) suggests that the western galaxy of the UGC~3841 pair (at a distance of 36 kpc from B2~0722+30) is most likely a dust-lane elliptical. This galaxy contains a weak, unresolved radio continuum source, MRK~1201, which has Seyfert 2 characteristics \citep{des00}. From our 1.4 GHz continuum data we derive a flux of $F_{\rm 1.4\ GHz} = 3.1$ mJy~beam$^{-1}$ ($P_{\rm 1.4\ GHz} = 2.3 \times 10^{21}$ W~Hz$^{-1}$) for this source. For clarification, we refer to this western galaxy of the UGC~3841 pair as MRK~1201. The eastern galaxy of this pair (which we will refer to as UGC~3841a) is an edge-on, distorted and elongated disc galaxy. The western part of the disc of UGC~3841a extends past MRK~1201 and points in the direction of B2~0722+30. MRK~1201 and UGC~3841a appear to be in heavy interaction, as can be seen from the kinematics of the \HI\ gas in Fig. \ref{fig:chanmaps0722}. The \HI\ gas follows the distorted disc of UGC~3841a, except in the velocity range $5629 \leq v \leq 5800$ \kms, where it is spread around MRK~1201. An \HI\ bridge, about 22 kpc in length, seems to stretch from the \HI\ distribution around MRK~1201 in the direction of B2~0722+30 ($5650 \leq v \leq 5714$ \kms\ in Fig. \ref{fig:chanmaps0722}). The total \HI\ mass detected in the MRK~1201/UGC~3841a pair is $3.4 \times 10^{9} M_{\odot}$.

To the north, at a distance of 33 kpc from UGC~3841a, is another \HI-rich galaxy pair. To our knowledge, this galaxy pair has not been classified previously. We refer to these galaxies as J072549+295830 (east) and J072548+295828 (west). This galaxy pair contains $4.2 \times 10^{8} M_{\odot}$ of \HI. 

\subsection{Stellar populations}
\label{sec:stelpop}

\begin{figure}
\centering
\includegraphics[width=0.48\textwidth]{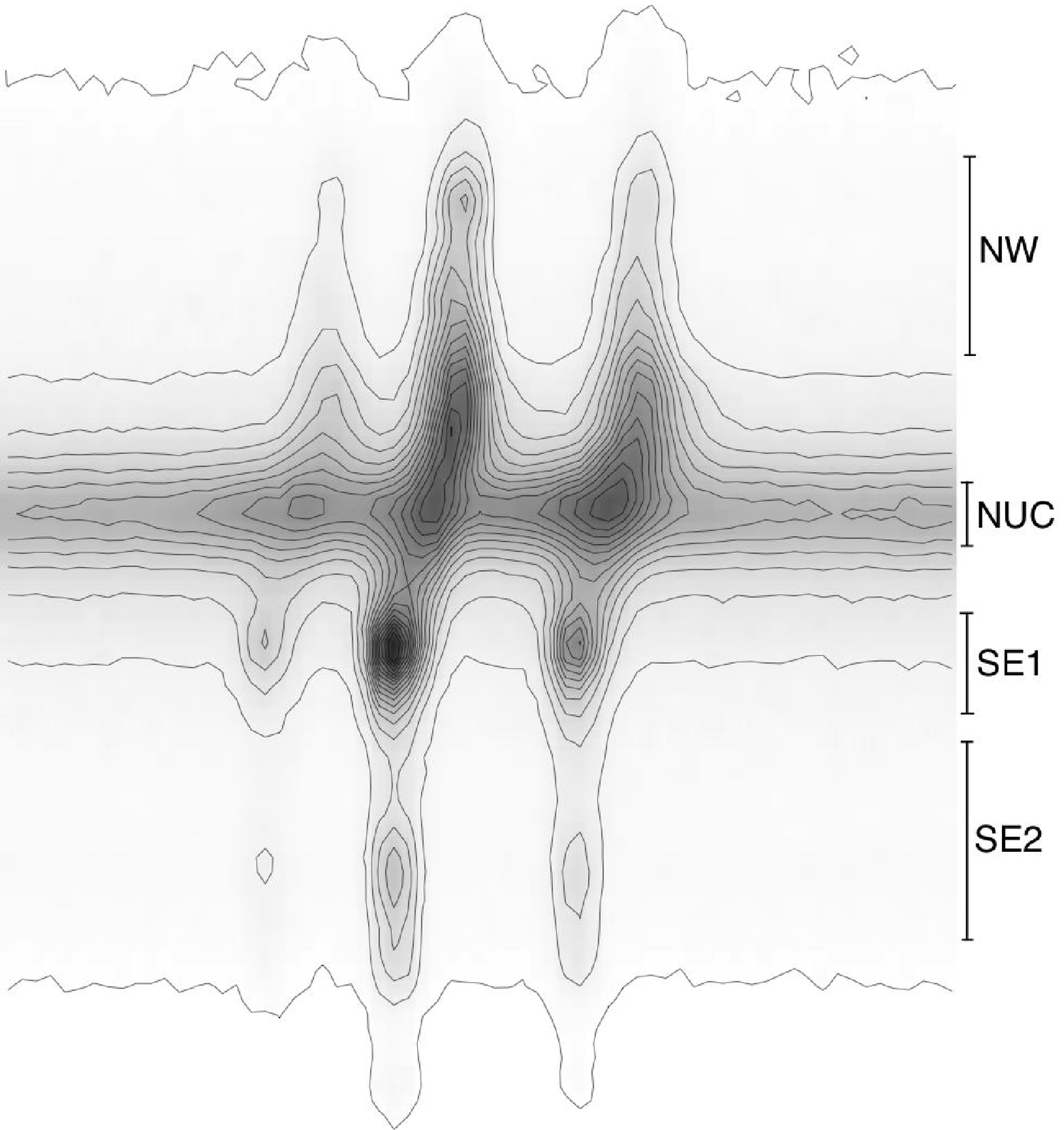}
\caption{2D-spectrum of B2~0722+30 around H$\alpha$+[\NII]$\lambda$$\lambda$6548,6583 (contour levels are between 10 and 90$\%$ of the maximum in steps of 5$\%$). On the right the various apertures that we used for our SED modelling are indicated.}
\label{fig:0722apertures}
\end{figure}

In order to study the stellar populations throughout B2~0722+30, we extracted spectra at various apertures along the optical emission-line disc. Figure \ref{fig:0722apertures} shows the location and Table \ref{tab:apertures} gives the details of these apertures. In Table \ref{tab:apertures}, columns 2 and 3 give the size of the aperture (in arcsec and kpc), while columns 4 and 5 give the distance of the centre of the apertures from the nucleus (assumed to be the brightest peak in the stellar continuum of the spectrum). The method we used for the stellar population analysis in the various regions is described in Sect. \ref{sec:SPmethod}.

\begin{table}
\centering
\caption{Apertures}
\label{tab:apertures}
\begin{tabular}{l|cc|cc}
      &  Width & Width & $D_{\rm nuc}$ & $D_{\rm nuc}$ \\
                &  (arcsec) & (kpc) & (arcsec) & (kpc) \\
\hline
NUC  &  2.0      & 0.8   &  -       &  -         \\  
SE1  &  3.2      & 1.2   &  4.2     &  1.6       \\  
SE2  &  6.0      & 2.3   &  9.2     &  3.6       \\  
NW   &  6.0      & 2.3   &  7.2     &  2.8       \\ 
\end{tabular} 
\end{table} 

\subsubsection{Nucleus}
\label{sec:nuc0722}

First we model the spectrum of the nuclear aperture of B2~0722+30. A single unreddened 12.5 Gyr OSP does not provide an accurate fit to the overall SED in the nuclear spectrum, i.e. the residuals of the model-fit to the spectrum are much larger than the flux calibration errors (Sect. \ref{sec:obsoptspec}). When including a power-law component in addition to the 12.5 Gyr OSP, a good fit can be obtained only for a power-law component with $\alpha = 2.8$, which contributes to about 50$\%$ of the light in the red part of the spectrum. A power-law with such a slope and high luminosity is not likely the result of an optical AGN in B2~0722+30, because there is no clear evidence for a bright point-source in the HST image of \citet{cap00} and because (as we will see in Sect. \ref{sec:emissionline}) the nuclear emission lines are characteristic of a Low Ionization Nuclear Emission-line Region (LINER) rather than an AGN power-law model. To investigate whether the power-law component could represent an additional reddening component that we did not take into account, we tried to fit a 12.5 Gyr OSP that we reddened by various amounts, but this did not produce a good fit.

\begin{figure*}  
\centering
\includegraphics[width=0.75\textwidth]{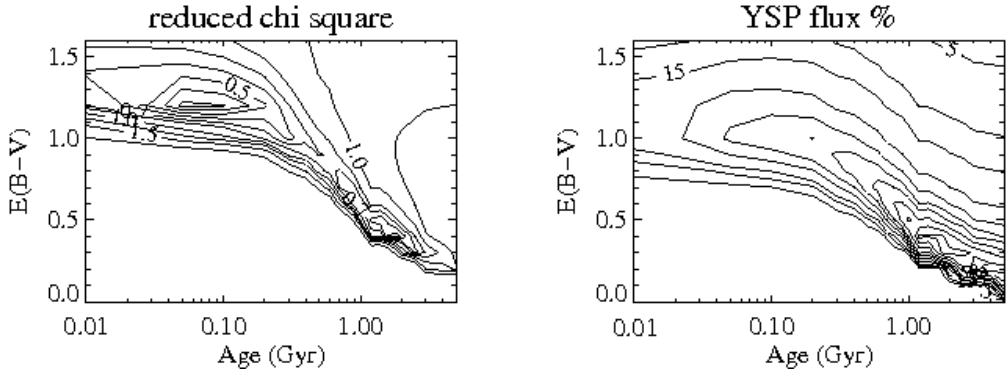}
\caption{\textbf{\textsl{B2~0722+30 NUC}} -- Least $\chi^{2}$ results of the SED modelling for different age and $E(B-V)$ of the YSP. Contour levels are: $\chi^{2}$ - 0.3, 0.4, 0.5, 0.6, 0.7, 0.8, 1.0, 1.2, 1.5, 2.0; Light$\%$ YSP - 5, 10, 15, 25, 35, 45, 55, 70, 85, 100.}
\label{fig:param0722}
\end{figure*}

\begin{figure*}
\centering
\includegraphics[width=0.64\textwidth]{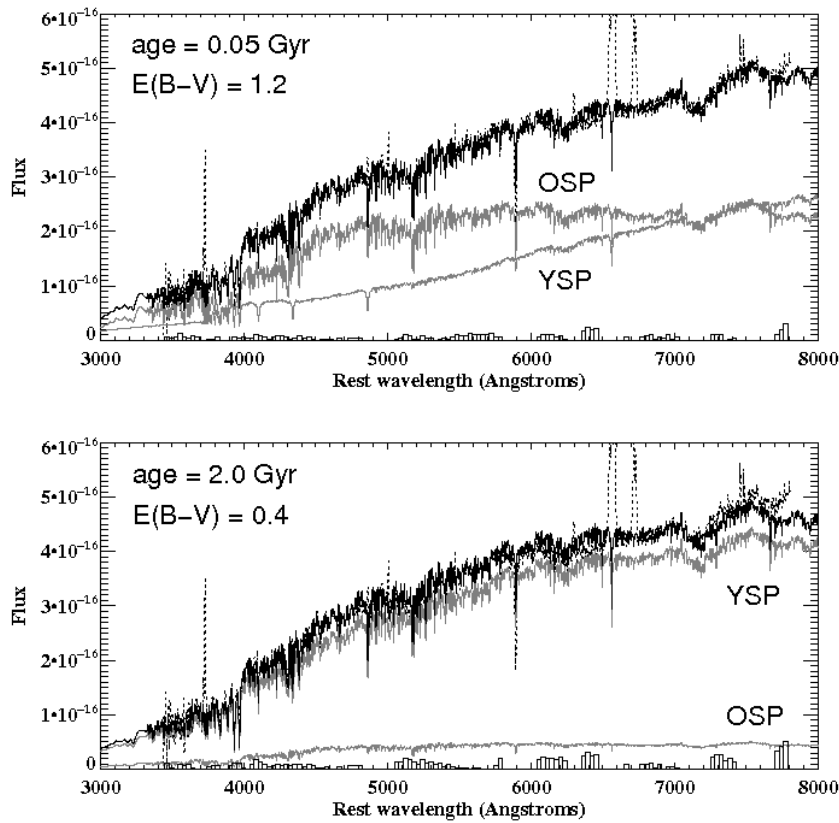}
\caption{\textbf{\textsl{B2~0722+30 NUC}} -- Least $\chi^{2}$ model fits to the nuclear spectrum of B2 0722+30. The dotted line is the observed spectrum, the black solid line is the fitted model and the grey solid lines are the separate components of the fitted model. The histogram at the bottom of the plots shows the residuals of the model fit to the observed spectrum. {\sl Top:} combination of a 12.5 Gyr old stellar population (OSP) and a young stellar population (YSP) of age = 0.05 Gyr and $E(B-V)$ = 1.2. {\sl Bottom:} combination of a 12.5 Gyr old stellar population (OSP) and an intermediate age stellar population of age = 2.0 Gyr and $E(B-V)$ = 0.4.}
\label{fig:bestfit0722nuc}
\end{figure*}

\begin{figure}
\centering
\includegraphics[width=0.49\textwidth]{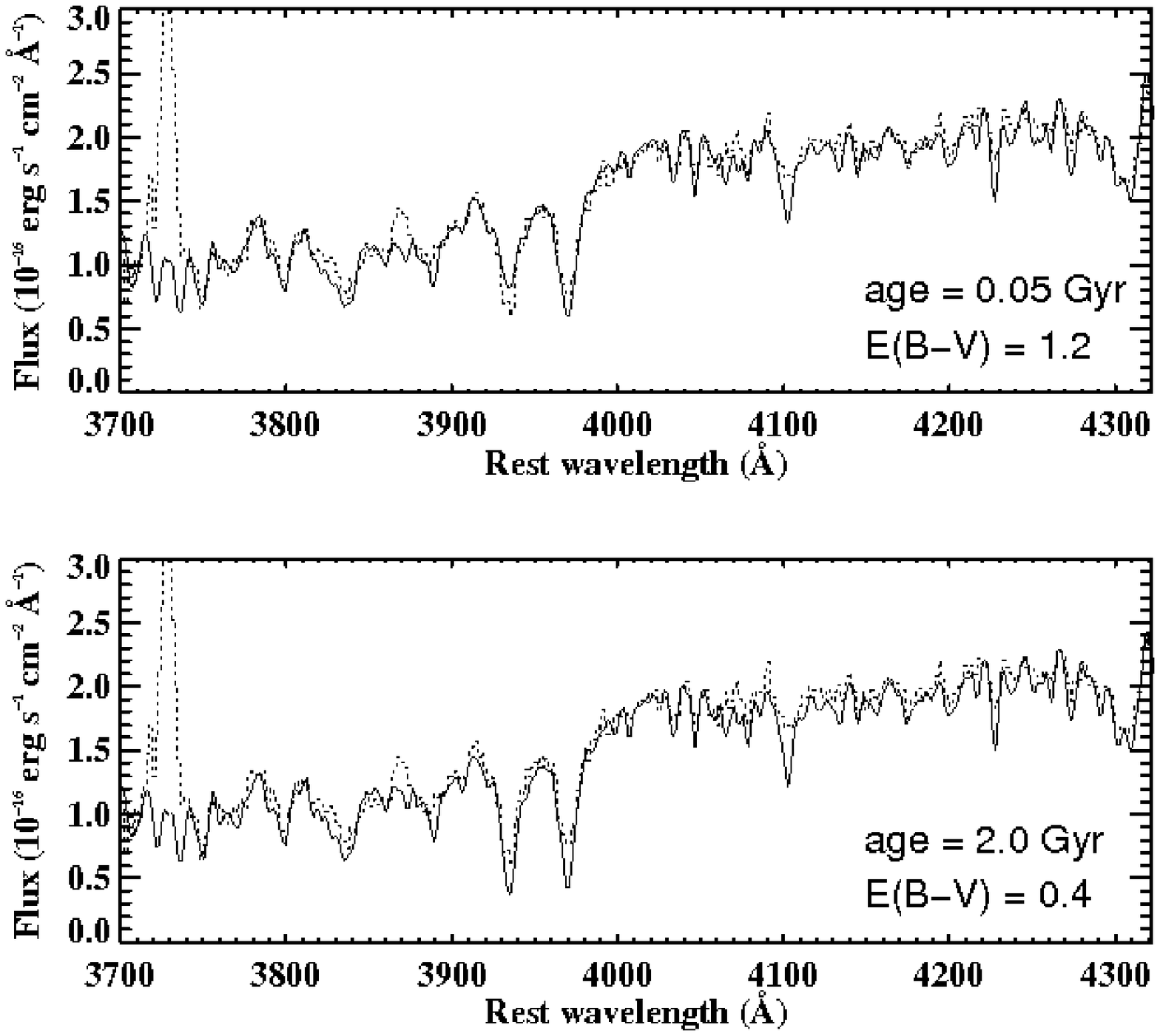}
\caption{\textbf{\textsl{B2~0722+30 NUC}} -- Least $\chi^{2}$ model fits to the nuclear spectrum of B2 0722+30 in the region around 4000 \AA. The dotted line is the observed spectrum, the black solid line the fitted model. {\sl Top:} combination of a 12.5 Gyr old stellar population (OSP) and a young stellar population (YSP) of age = 0.05 Gyr and $E(B-V)$ = 1.2. {\sl Bottom:} combination of a 12.5 Gyr old stellar population (OSP) and an intermediate age stellar population of age = 2.0 Gyr and $E(B-V)$ = 0.4.}
\label{fig:bestfit0722nuczoom}
\end{figure}

Since no good or physically acceptable fit of a single 12.5 Gyr OSP or OSP + power-law component could be obtained for the nuclear region of B2~0722+30, we subsequently modelled the nuclear spectrum with a combination of a 12.5 Gyr OSP and a (reddened) YSP. Figure \ref{fig:param0722} shows the results of adding a YSP. The solution of fitting an OSP + YSP clearly converges to two different models: including a YSP of age 0.05 Gyr, $E(B-V) = 1.2$ and light percentage = 30$\%$ results in reduced $\chi^{2} = 0.27$, the least $\chi^{2}$ fit across our range of parameter space; including a YSP of age 2.0 Gyr, $E(B-V) = 0.4$ and light percentage = 89$\%$ gives a least $\chi^{2}$ solution of 0.52 in this part of the parameter space. These two model fits are shown in Fig. \ref{fig:bestfit0722nuc}. In an attempt to break this degeneracy, we look in detail at the age sensitive absorption features around 4000 \AA. Figure \ref{fig:bestfit0722nuczoom} (top) shows that both models provide a reasonable fit to the age sensitive diagnostic Balmer and \Cak($\lambda$3933)/\Cah($\lambda$3968) absorption lines and the G-band feature around $\lambda$ 4305 \AA. Overall, the 0.05 Gyr YSP seems to provide the best fit to most of the features, except for \Cak. Narrow \Cak\ absorption by the interstellar medium (ISM) might affect the line-strength in the core of the stellar \Cak\ line \citep[e.g.][]{pas00}, resulting in a deeper \Cak\ absorption feature in the observed spectra compared with our model. The fit of an intermediate age stellar population with moderate reddening (age 2.0 Gyr and $E(B-V) = 0.4$) provides a better relative fit to both calcium lines (Figure \ref{fig:bestfit0722nuczoom} - bottom), but appears to slightly over predict the depth of the lines. We note, however, that part of this could be due to a slight difference in resolution between the real and the template spectra. We also tried to fit a two-component combination of a 2 Gyr OSP plus 0.05 Gyr YSP to the nuclear region, but due to the dominating contribution of the 2.0 Gyr stellar population in that case ($>$85$\%$ of the total stellar light) this does not provide accurate information about a possible YSP component. Including an additional power-law component to our model fits did not significantly improve or alter our results.

We therefore argue that our used method cannot clearly distinguish between a dominating 2.0 Gyr stellar population or a combination of a 12.5 Gyr OSP and a heavily reddened 0.05 Gyr YSP in the nuclear region of B2~0722+30 (which will be dominated by the bulge component). Further limitations of our used method are discussed in Sect. \ref{sec:uncertainties}.

\subsubsection{Off-nuclear regions}
\label{sec:0722offnuc}

The spectra in the off-nuclear regions (SE1, SE2 and NW) cannot be fitted accurately with a single unreddened 12.5 Gyr OSP and an AGN power-law component is not likely to be important far from the nucleus (which did not need the inclusion of a power-law component anyway).

\begin{figure*}
\centering
\includegraphics[width=\textwidth]{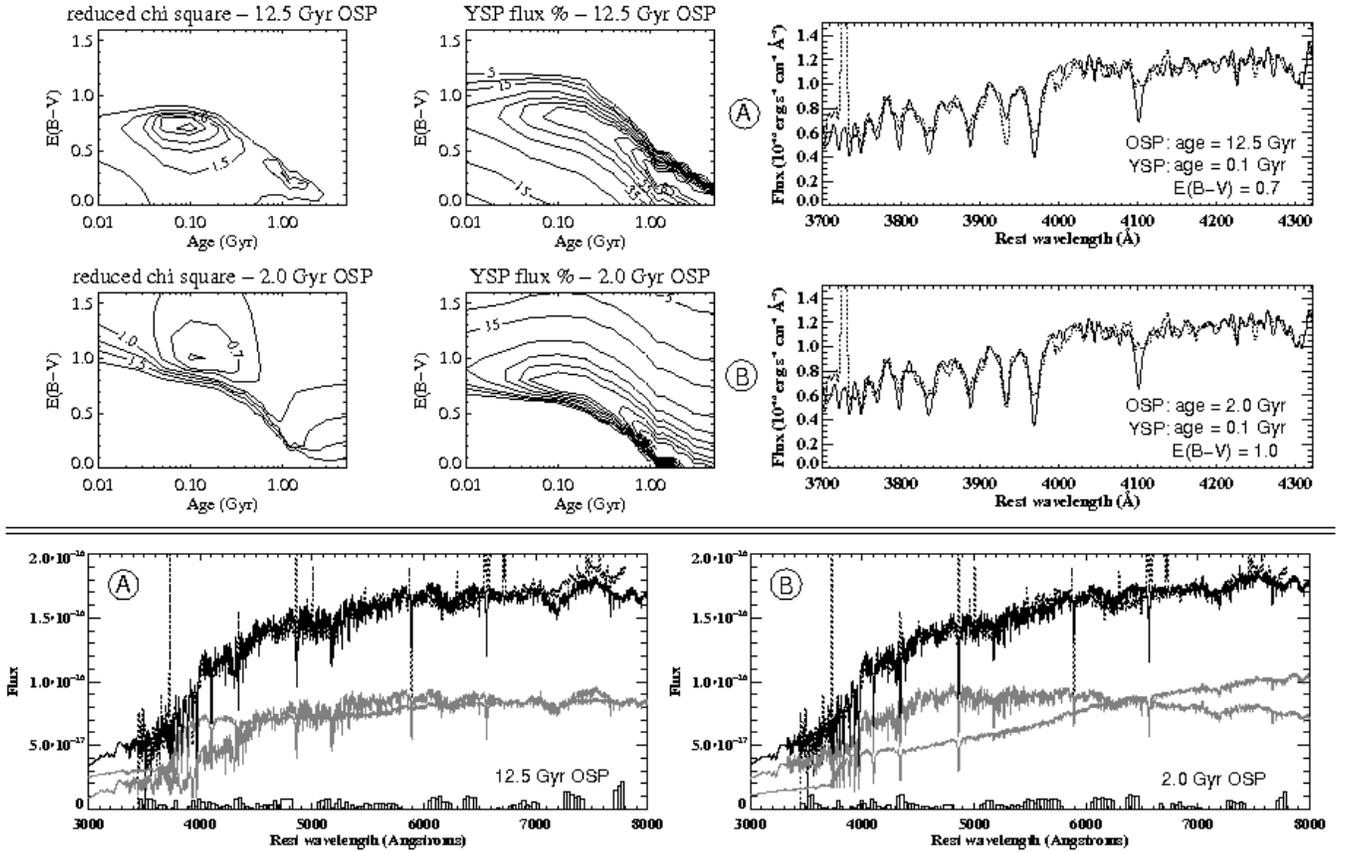}
\caption{\textbf{\textsl{B2~0722+30 SE1}} -- Model 'A' includes an unreddened OSP of 12.5 Gyr; model 'B'  an unreddened OSP of 2.0 Gyr. {\sl Top left:} least $\chi^{2}$ results of the SED modelling for different age and $E(B-V)$ of the YSP. Contour levels are the same as in Fig. \ref{fig:param0722}. {\sl Top right:} Detailed least $\chi^{2}$ model fit in the region around 4000 \AA. The dotted line is the observed spectrum, the black solid line the fitted model. {\sl Bottom:} Least $\chi^{2}$ model fit to the entire SED. The dotted line is the observed spectrum, the black solid line the fitted model and the grey solid lines the separate components of the fitted model. The histogram at the bottom of the plots shows the residuals of the model fit to the observed spectrum.}
\label{fig:0722_bestfit_SE1_outer}
\end{figure*}

Figure \ref{fig:0722_bestfit_SE1_outer} shows the results of adding a YSP in fitting the spectrum of region SE1. As can be seen from 'case~A', when adding a YSP of age $\sim$ 0.05 -0.1 Gyr and $E(B-V) \approx 0.4 - 0.7$ to the 12.5 Gyr OSP, we get a reasonable fit to the SED in the region SE1. However, when looking in detail at the model fit of case~A (12.5 Gyr OSP + YSP), we see that the fit is not very accurate to the age-sensitive \Cak\ line. Although narrow \Cak\ absorption by the ISM might affect the line strength in the core of the stellar \Cak\ line \citep[e.g.][]{pas00}, the asymmetry between the observed spectrum and our model fit for the \Cak\ line in the off-nuclear region SE1 is much larger than for the nuclear region. We do not expect that this can be explained solely by \Cak\ absorption by the ISM. This questions the validity of our SED fitting results in region SE1. However, we saw that the nuclear region may be dominated by an intermediate age stellar population of 2.0 Gyr instead of a 12.5 Gyr OSP. We therefore also modelled our off-nuclear spectrum in region SE1 with an unreddened 2.0 Gyr ``OSP'' combined with a YSP ('case~B' in Fig. \ref{fig:0722_bestfit_SE1_outer}). Substituting an underlying 12.5 Gyr OSP for a 2.0 Gyr OSP gives a much better fit to the \Cak\ line. At the same time, the least $\chi^{2}$ solution for the overall SED fit does not change much. For the case~B modelling the YSP has an age of about 0.05 (NW) - 0.1 (SE1 and SE2) Gyr and contributes to about 35-55$\%$ of the total stellar light. The reddening is slightly higher than for the case~A modelling. 

\begin{figure*}
\centering
\includegraphics[width=0.95\textwidth]{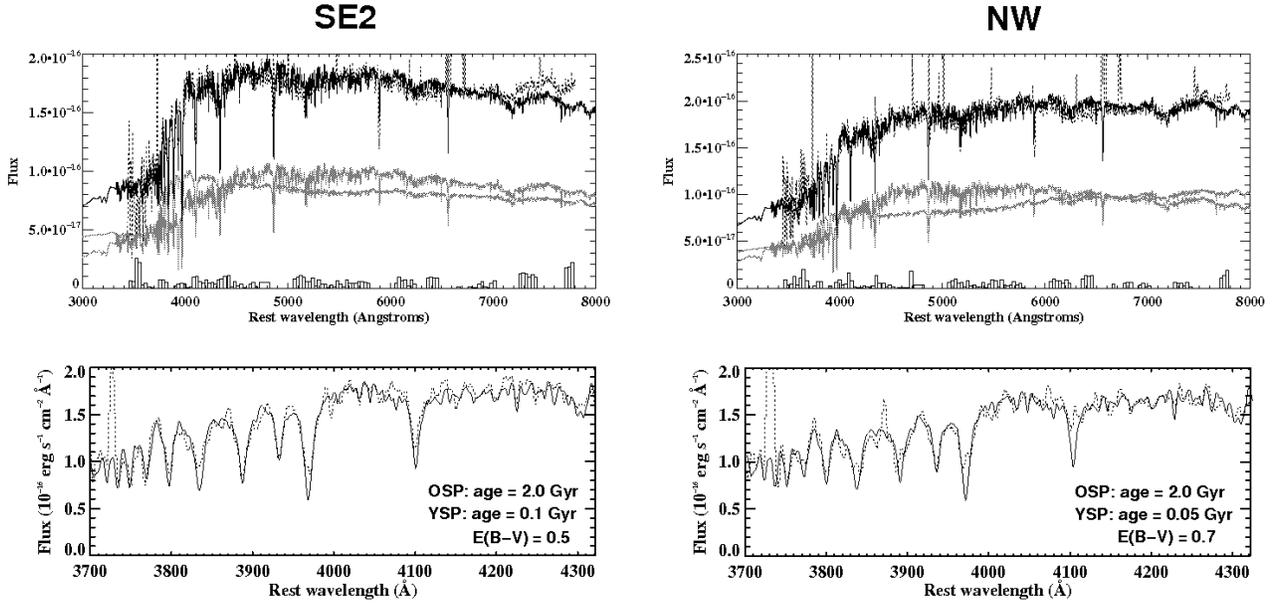}
\caption{\textbf{\textsl{B2~0722+30 SE2 and NW}} Best fit of a 2.0 Gyr OSP + reddened YSP to the spectra in regions SE2 and NW. {\sl Top:} Least $\chi^{2}$ model fit to the entire SED. The dotted line is the observed spectrum, the black solid line the fitted model and the grey solid lines the separate components of the fitted model. The histogram at the bottom of the plots shows the residuals of the model fit to the observed spectrum. {\sl Bottom:} Detailed least $\chi^{2}$ model fit in the region around 4000 \AA. The dotted line is the observed spectrum, the black solid line the fitted model.}
\label{fig:0722_bestfit_SE2NW_outer}
\end{figure*}

The story for regions SE2 and NW is very similar to that of region SE1. Also for these regions, the combination of a 12.5 Gyr OSP + YSP gives an acceptable least $\chi^{2}$ solution, but again severely under-predicts the strength of the \Cak\ line. Substituting the 12.5 Gyr OSP for a 2.0 Gyr OSP provides a good fit to the age-sensitive \Cak-line, as well as the Balmer absorption lines and G-band feature, without changing the age and light contribution of the YSP dramatically. The best $\chi^{2}$ solution and fit to the spectra in regions SE2 and NW are shown in Fig. \ref{fig:0722_bestfit_SE2NW_outer}. 

Therefore -- although there is a substantial uncertainty in the properties of the OSP across B2~0722+30 as well as the stellar content of the bulge region -- the presence of a YSP across the disc of B2~0722+30 is fairly well established with our stellar population method. Table \ref{tab:modelparam} summarises the best-fit results from our stellar population analysis of B2 0722+30. The light percentage coming from the YSP is roughly 40-50$\%$ for all the three off-nuclear regions. This strongly suggests that the stellar population is fairly uniform across the disc of B2~0722+30. It is interesting to note that the reddening of the YSP, $E(B-V)$, is highest in the inner region SE1 ($E(B-V) \approx 1.0$) and gets lower for the outer regions ($E(B-V) \approx 0.7$ in NW and $\approx 0.5$ in SE2).


\subsubsection{Stellar masses}
\label{sec:stellarmasses}

The masses of the OSP and YSP in the off-nuclear apertures are also given in Table \ref{tab:modelparam}. The masses are calculated from the flux of the YSP and OSP at 4770 \AA\ (the central $\lambda$ of our normalising bin). The YSP and OSP fluxes (taking into account reddening; see Table \ref{tab:modelparam}) are scaled to stellar masses using the template spectra of \citet{bru03}. The uncertainty in the stellar mass estimates is based on uncertainties from the SED modelling and does not include the uncertainty of a factor of 2 due to slit-losses during the observations (see Sect. \ref{sec:obsoptspec} for details).


\subsubsection{Infrared properties}
\label{sec:IR}

B2~0722+30 has a far-infrared (FIR) luminosity of $L_{\rm FIR} = 4.7 \times 10^{10} L_{\odot}$ \citep[][corrected for H$_{0} = 71$ \kmsMp]{maz93}, similar to that of star forming Sb and Sc galaxies \citep{rie86}. This is a relatively high FIR luminosity, given that -- as we will see in Sect. \ref{sec:spiralhost} -- B2~0722+30 is relatively small compared with for example the Milky Way galaxy. In order to investigate whether dust-heating by the YSP in B2~0722+30 could be responsible for this high FIR luminosity, we calculate the total bolometric luminosity ($L_{\rm bol}$) of the 0.05-0.1 Gyr YSP across B2~0722+30. $L_{\rm bol}$ is estimated from the integrated luminosity of the unreddened YSP measured from our template spectra (across the wavelength range 0 - 30,000 \AA), scaled to the total mass of the YSP given in Table \ref{tab:modelparam}. Given that the reddening is $E(B-V) = 0.5 - 1.0$ across the disc, it is likely that most ($\ga 85\%$) of the total bolometric luminosity of the young stars is absorbed by dust. Assuming that most of this dust-absorbed light is re-radiated into the FIR, $L_{\rm bol}$ should therefore provide a reasonable estimate for the FIR luminosity related to the YSP in B2~0722+30. 

We estimate that $L_{\rm bol}$ of the YSP in B2~0722+30 is between $0.3 \times 10^{10} L_{\odot}$ and $1.5 \times 10^{10} L_{\odot}$, depending on the age of the YSP varying between 0.05 and 0.1 Gyr and whether or not a 0.05 Gyr YSP in the nuclear region is taken into account (see Table \ref{tab:modelparam}). We argue that these estimates are likely lower limits, given that our apertures do not cover the entire optical extent of the host galaxy. In addition, stars which formed less than $0.05-0.1$ Gyr ago (which are not identified with our method for modelling the stellar population) provide the highest relative contribution to the YSP's bolometric luminosity. We note, however, that this estimate does not include the factor 2 uncertainty in absolute flux calibration of our spectra (see Sect. \ref{sec:obsoptspec})

In order to circumvent the limited slit coverage and uncertainty in absolute flux calibration and to obtain a better estimate of the total bolometric luminosity of the entire YSP in B2~0722+30, we apply the average values of our SED modelling to the total V-band luminosity of the system. \citet{gon00} measured the optical V-band magnitude of the host galaxy B2~0722+30 to be $m_{V} = 15.04$, which corresponds to $L_{V} = 5.0 \times 10^{9} L_{\odot}$. Assuming that 43$\%$ of the V-band light across B2~0722+30 comes from a 0.05-0.1 Gyr YSP with reddening $E(B-V) = 0.73$, we estimate that $L_{\rm bol} = 3.5 - 5.0 \times 10^{10} L_{\odot}$, which is in excellent agreement with $L_{\rm FIR}$ measured from the IRAS data. We therefore argue that the total $L_{\rm bol}$ of the YSP appears to be roughly consistent with $L_{\rm FIR}$ in B2~0722+30 and that dust heating by the YSP in B2~0722+30 is likely responsible for at least a substantial fraction of the relatively high FIR luminosity of the host galaxy.

\subsubsection{Uncertainties in our method}
\label{sec:uncertainties}

\begin{table*}
\caption{Best results of the least $\chi^{2}$ model fitting}
\centering
\begin{tabular}{l|ccc|cc|cc}
Aperture     &  Age OSP & Age YSP & $E(B-V)$ & Light $\%$ & Mass $\%$ & Mass ($M_{\odot}$) & Mass ($M_{\odot}$) \\
             &  (Gyr)   & (Gyr)   & YSP      &    YSP    &   YSP  & YSP & OSP \\ 
\hline
 NUC (1) & 12.5 & 0.05            & 1.2        & 30      &  22         & $4.1 \times 10^{8}$       & $3.4 \times 10^{9}$ \\ 
 NUC (2) & 12.5 & 2.0             & 0.4        & 89      &  83         & $2.4 \times 10^{9}$       & $5.3 \times 10^{8}$ \\
 SE1  & 2.0  & 0.1                & 1.0        & 37      &  61         & $2.1 \times 10^{8}$       & $2.7 \times 10^{8}$ \\  
 SE2  & 2.0  & 0.1                & 0.5        & 48      &  29         & $8.4 \times 10^{7}$       & $2.8 \times 10^{8}$  \\ 
 NW   & 2.0  & 0.05               & 0.7        & 44      &  32         & $8.9 \times 10^{7}$       & $3.0 \times 10^{8}$ \\ 
\end{tabular}
\flushleft
Notes -- The light$\%$ is the relative flux contribution of the YSP in the normalising bin (4720 - 4820 \AA). The mass of the stellar populations in the various apertures (last two columns) is derived from the stellar population's light$\%$ of the observed flux at 4800 \AA\ - corrected for reddening - and scaled to the unreddened model template spectra.
\label{tab:modelparam}
\end{table*} 

Of course there is a degree of uncertainty introduced by observational errors (such as flux calibration), the uniqueness of the model solutions (see Sect. \ref{sec:nuc0722} and \ref{sec:0722offnuc}) and by the assumed parameters of the synthesis models that we used in our fitting procedure (duration of the starburst, the assumed Salpeter IMF and  metallicity).

Uncertainties in the flux calibration could be the reason that in particular in the red part of the spectrum the fit is not perfect. However, we do not expect that this will significantly change our results, since our visual inspection of the detailed fits in the blue part of the spectrum (around 4000 \AA) is in excellent agreement with the reduced $\chi^{2}$ results. There is also the selection effect that the apertures that we chose for our study cover regions of a kpc or more in various parts of the galaxy's disc. Locally, the stellar populations and dust properties could vary from place to place within these apertures. Nevertheless, the similarity of the results for the various off-nuclear regions reassures us that outside the nucleus the average values for the stellar ages and masses are fairly uniform. All four apertures combined do not completely cover the total extent of the galaxy, therefore the total mass of the stellar populations in B2~0722+30 is somewhat larger than the sum of the stellar masses across the separate apertures.

For a detailed discussion of the uncertainties in the assumed parameters of the synthesis models we refer to \citet{tad02}. As discussed there, the major uncertainty in the model parameters is the assumed shape of the IMF. This may lead to uncertainties of a factor of 2-3 in the total mass estimates in Table \ref{tab:modelparam} (on top of the factor of 2 due to observational inaccuracies; Sect. \ref{sec:obsoptspec}). Regarding the metallicity, we only used template spectra with solar-abundances to fit our data. Following for example \citet{bla06}, uncertainty in the metallicity could cause a serious scatter in the derived Single Stellar Population (SSP)-equivalent age of a stellar population. However, we do not expect that metallicity effects significantly influence the age and mass estimates that we derive for the YSP detected in the disc of B2~0722+30, where one expects solar metallicities. \citet{sar05} investigated the effect of metallicity on an SED fitting technique for the bulge-regions of nearby galaxies. They show that substituting solar for 2.5 $\times$ solar metallicity YSP template spectra does not change the results for age and light percentage of the YSP dramatically. The most significant effect on the YSP is that super-solar metallicities can mimic additional reddening, in which case our solar-abundance models might overestimate the derived $E(B-V)$. Metallicity effects could, however, seriously affect the results for the older 2.0 Gyr stellar population that we fit throughout B2~0722+30 and which may dominate the stellar mass in the bulge region. It is known that there is a broad range in metallicities across galaxy bulges \citep[e.g.][]{wys97} and that metallicities in the central bulge can be significantly higher than in the disc. Moreover, for old and intermediate age stellar populations such as the older stellar population in B2~0722+30 an age-metallicity degeneracy is likely to be much more significant than for young stellar populations. Therefore, our results for the 2.0 Gyr stellar population in B2~0722+30 should be taken with care and we warn the reader that the estimated age of this older stellar population should not be take at face-value. Finally, our used models are based on fitting a single, instantaneous starburst event on top of an old stellar population. More subtle, previous starbursts or longer periods of continuous star formation cannot be detected with our method. It is very likely that the YSP detected in the disc of B2~0722+30 is evidence for {\sl ongoing} star formation over a substantial period, as is normal in spiral galaxies.

\subsubsection{Stellar populations in B2~0722+30 (summary)}
\label{storyB20722}

The analysis of the spectra of B2 0722+30 shows that a 0.05 - 0.1 Gyr young stellar population contributes to about half of the stellar light in the disc of B2~0722+30. This YSP most likely traces ongoing star formation (as we will see in Sect. \ref{sec:emissionline}, in the disc of B2~0722+30 the emission-line ratios resemble those of star-forming regions). We argue that there is a significant uncertainty only in the total mass and perhaps in the reddening of this young stellar population. The stellar content of the bulge region as well as the properties of the underlying old stellar population throughout the galaxy are much more uncertain and could not be properly constrained with our used SED modelling technique.

\subsection{Emission-line diagnostics}
\label{sec:emissionline}

\begin{table*}
\centering
\caption{Line ratios}
\label{tab:lineratios}
\begin{tabular}{l|cccccc|l}
Aperture         & ${{[\rm \OIII]}\over{H_{\beta}}}$ & ${{[\rm \OII]}\over{H_{\beta}}}$ & ${{[\rm \OII]}\over{[\rm \OIII]}}$ & ${{[\rm \NII]}\over{H_{\alpha}}}$ & ${{[\rm \SII]}\over{H_{\alpha}}}$ & ${{[\rm \OI]}\over{H_{\alpha}}}$ & type \\
\hline
NUC &  1.5  & 3.3 & 2.2 & 1.2 & 0.8 & 0.14 & LINER \\
SE1 &  0.3  & 0.9 & 2.6 & 0.5 & 0.3 & 0.04 & \HII\ region \\
SE2 &  0.4  & 1.3 & 3.2 & 0.5 & 0.3 & 0.06 & \HII~/\ LINER \\
NW  &  0.8  & 1.7 & 2.1 & 0.6 & 0.4 & 0.05 & \HII~/\ LINER \\
\end{tabular} 
\flushleft
{Notes -- Line-ratios are given for the emission lines \Hb$\lambda4861$, \Ha$\lambda6563$, [\OIII]$\lambda5007$, [\OII]$\lambda3727$, [\NII]$\lambda6583$, [\SII]$\lambda6716$+$\lambda6731$ and [\OI]$\lambda6300$. The last row gives the characteristic type of emission-line object that corresponds to the line-ratios \citep[as given by][]{bal81,vei87}.}
\end{table*} 

In order to investigate the optical properties of the AGN and to get an idea about the ionisation-mechanism of the emission-line gas in B2~0722+30, we measured the emission-line ratios in the nuclear as well as the off-nuclear regions. Emission-line infilling by stellar-related absorption features (in particular in the hydrogen lines) could be significant, therefore we subtracted the best-fit model from our observed spectra to eliminate the underlying stellar continuum and absorption as well as possible. Table \ref{tab:lineratios} gives the emission-line ratios in the various apertures. The last column gives the corresponding characteristic spectral classification based on diagnostic diagrams of \citet{vei87} and \citet{bal81}. In particular the diagnostic diagrams of \citet{vei87} are based on the ratios of emission lines that are close together in order to minimize the effects of reddening. The [\OIII]$\lambda$5007 luminosity that we measure in the nuclear aperture (corrected for a galactic extinction of $E(B-V) = 0.059$; see Sect. \ref{sec:obsoptspec}) is $L_{\rm [\OIII]} = 1.3 \times 10^{39}$ erg~s$^{-1}$.

The nuclear region shows the emission-line characteristics of a Low Ionisation Nuclear Emission-line Region (LINER). The LINER classification for B2~0722+30 can clearly be distinguished from a Seyfert-type nucleus in the diagnostic diagrams of \citet[][]{vei87} and \citet[][]{bal81}. We note, however, that our nuclear aperture covers a region of about 0.8 kpc, which is significantly larger than the central black-hole region. Nevertheless, a bright Seyfert nucleus would likely dominate the emission-line flux in this nuclear aperture and no indications of such a strong optical AGN are present in our spectra. The emission-line spectrum of the nuclear region therefore indicates that the optical AGN has the characteristics of a low-excitation AGN, as is common for \FRI\ radio galaxies. We will discuss this further in Sect. \ref{sec:AGNclas}.

The off nuclear regions in the disc of B2 0722+30 show emission-line characteristics that resemble those of \HII\ regions. This strengthens the idea that the YSP detected across the disc of B2~0722+30 represents ongoing star formation.

\section{Discussion}
\label{sec:discussion}

In this Section, we discuss in detail the properties of the host galaxy and AGN in B2~0722+30. We compare the characteristics of B2~0722+30 with two other disc-dominated radio galaxies in order to investigate in more detail the circumstances under which a double-lobed radio source can occur in these disc-galaxies.

\subsection{Disc-dominated host galaxy}
\label{sec:spiralhost}

The optical morphology, gas content/distribution and stellar population analysis described in this paper clearly show that B2~0722+30 is a galaxy with a prominent star-forming disc. The optical morphology of B2~0722+30 is most clearly seen in the ${\sl HST}$ image of \citet{cap00}, who classify the galaxy as a spiral. However, given the large inclination of the galaxy, direct evidence for a spiral morphology in the form of prominent spiral arms is lacking and hence the classification of B2~0722+30 is often ambiguous in the literature (spiral or S0). From V-band surface photometry, \citet{gon00} classify B2~0722+30 as a spiral or S0 galaxy and conclude that its radial-light profile has discy isophotes that imply the presence of either a disc, arms or a bar.

\begin{figure}
\centering
\includegraphics[width=0.45\textwidth]{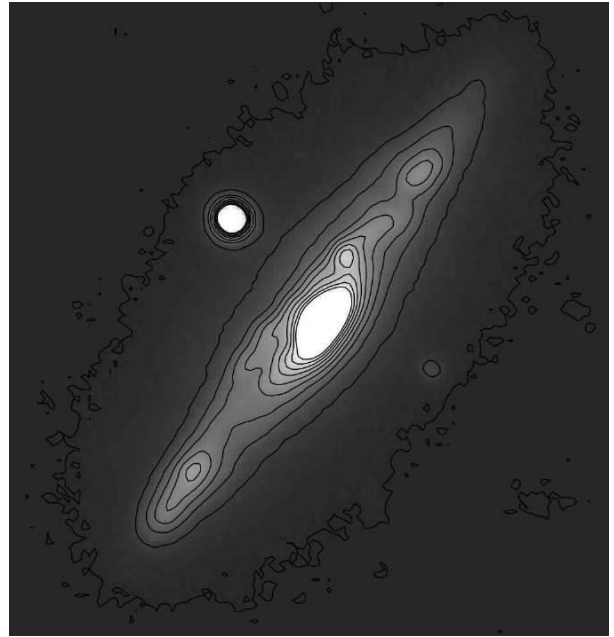}
\caption{Deep optical B-band image of B2~0722+30. Contours are in arbitrary count-units and range from 5 to 900 in steps of 100. The innermost contour marks the bulge region, based on visual comparison with the HST image of \citet{cap00}.}
\label{fig:bulge}
\end{figure}

Figure \ref{fig:bulge} shows in more detail our deep optical B-band image of B2~0722+30. There is clear morphological structure visible along the stellar disc. A comparison with the ${\sl HST}$ image of \citet{cap00} shows that at least part of this structure is caused by the presence of large amounts of dust across the disc, although the exact extent of the dust remains somewhat ambiguous from our data (as well as from the limited field-of-view of the ${\sl HST}$ image). The presence of a gas-rich disc with large-scale dust and active star formation fits the classical definition of a late-type system, although we note that similar properties have recently been observed for NGC~612 \citep{hol07,emo08}, which was classified as an archetypal S0 from its optical morphology \citep{ver01}. Figure \ref{fig:bulge} also shows that the bulge region of B2~0722+30 (marked by visually comparing our B-band image with the ${\sl HST}$ image of Capetti et al.) is modest compared to the total extent of the disc. We derive an estimated bulge-to-disc flux ratio of B/D $\sim$ 0.2 from our B-band image (which is likely an upper limit, given the large inclination and dust-obscuration of the disc). This B/D is in agreement with what is generally found for spiral galaxies \citep[although the average B/D for S0s is only slightly higher than our estimated upper limit;][]{gra08}. 

From the observational evidence, we therefore conclude that B2~0722+30 is a gas-rich disc galaxy. Although its characteristics appear to fit the description of a spiral galaxy, the exact morphological classification is not expected to influence the main conclusions of this paper, given that for for disc galaxies in general (both spirals and S0s) it is extremely rare to host a classical radio source (see Sect. \ref{sec:introduction} for details).\\
\vspace{0mm}\\
Our \HI\ and optical emission-line data show that B2~0722+30 contains a regularly rotating disc of both cool neutral hydrogen gas and ionised emission-line gas. The total mass of neutral hydrogen gas in the disc ($M_{\rm HI} = 2.3 \times 10^{8} M_{\odot}$, or a somewhat higher value when taking into account the significant amounts of \HI\ gas detected in absorption against the radio continuum) is a factor of five lower than the mass of molecular hydrogen in B2~0722+30 \citep[$M_{\rm H2} = 1.2 \times 10^{9} M_{\odot}$, as derived from single-dish CO observations by][]{eva05}, and about an order of magnitude lower than the \HI\ mass of our Milky Way galaxy \citep[$M_{\rm HI\ MW} = 5 \times 10^{9} M_{\odot}$;][]{hen82}. Also, the total extent of the \HI\ disc in B2~0722+30 ($\sim 15$ kpc) is less than half the diameter of the Milky Way \HI\ disc. As mentioned in Sect. \ref{sec:IR}, the V-band luminosity of B2~0722+30 is $L_{V} = 5.0 \times 10^{9} L_{\odot}$ \citep{gon00}. Therefore, $M_{\rm HI}/L_{V} = 0.05$. Albeit at the low end, this is consistent for what is observed in spiral galaxies \citep[e.g][please note that our derived value of $M_{\rm HI}/L_{\rm V}$ is somewhat larger when taking into account the \HI\ gas detected in absorption]{kna89}. The average surface density of the disc is about $4.1\ M_{\odot}\ {\rm pc^{-2}}$, which is close to the critical gas surface density for star formation in galaxy discs \citep{mar01,hul93}. The YSP across the disc and the emission-line ratios characteristic of \HII\ regions in the ionised gas disc indicate that star formation is happening across the disc of B2~0722+30.

Given the regular rotation of the \HI\ gas in the disc of B2~0722+30, we can make an estimate for the total mass enclosed by the system by assuming that the underlying dark matter halo has a spherical distribution. In that case, 
\begin{equation}
M_{\rm enc} = {{R_{\rm out} \times v_{\rm out}^{2}}\over{{\rm sin}^{2}i\ G}} = 8.8 \times 10^{10}\ {\rm sin}^{-2}i\ M_{\odot}, 
\end{equation}
with $R_{\rm out} = 7.5$ kpc the outer radius of the rotating \HI\ disc, $v_{\rm out} = 225$ \kms\ the observed velocity of the disc at this distance, $i$ the inclination of the disc (close to 90$^{\circ}$ for B2~0722+30) and $G = 6.673 \times 10^{-11}$ m$^{3}$ kg$^{-1}$ s$^{-2}$ the gravitational constant.

\subsection{AGN classification}
\label{sec:AGNclas}

A detailed continuum image of the radio source in B2~0722+30 is presented by \citet[][also shown in Fig. \ref{fig:jetalignment}]{fan86}. From the continuum emission in our 21cm data we derive a total radio power of $P_{\rm 1.4GHz} = 1.1 \times 10^{23}$ W~Hz$^{-1}$ for B2~0722+30, similar to the value found by \citet{fan86}. The optical V-band magnitude of B2~0722+30 is $m_{V} = 15.04$ \citep{gon00}. Following \citet{oke74}, this corresponds to an optical flux density of $3.5 \times 10^{-26}$ ergs~s$^{-1}$~cm$^{-2}$~Hz$^{-1}$. The radio-to-optical flux ratio of B2~0722+30 is therefore $R_{\rm 1.4~GHz} = 41$. \citet{kel89} defined $R_{\rm 5~GHz} = 10$ as the border between radio-quiet and radio-loud objects, which would correspond to roughly $R \sim 19$ at 1.4~GHz \citep[see][]{kom06}. B2~0722+30 is therefore clearly a radio-loud object, although its total radio power lies in the transition region between that of Seyferts and radio galaxies. Radio-loud Seyfert galaxies with total radio power and radio-to-optical flux ratio similar to that of B2~0722+30 do exist, but they are rare and their radio component is generally very compact \citep[e.g.][and references therein]{kom06}. An extreme example is the nearby Seyfert 2 galaxy NGC~1068 \citep[e.g.][]{wil87}, which has a 1 kpc-scale radio jet with a total radio power of the same order as B2~0722+30. In contrast to Seyfert galaxies, however, the radio source in B2~0722+30 has a diameter of about 13.6 kpc \citep{fan87} and stretches outside the optical body of the host galaxy. Although its linear size is still rather small compared to the majority of classical double-lobed radio sources, B2~0722+30 clearly contains a two-sided jet structure \citep[resembling so called 'naked jets', as discussed in detail by][]{par87}. The morphology of this two-sided jet extending over many kpc classifies B2~0722+30 as an \FRI\ radio source.

In Sect. \ref{sec:emissionline} we derived the emission-line ratios in the nuclear region of B2~0722+30, which provide further insight in the properties of the AGN. The characteristic LINER classification of the optical emission-line AGN is also in agreement with that of low-excitation \FRI\ radio galaxies. As already explained in Sect. \ref{sec:emissionline}, the emission-line ratios are inconsistent with a typical Seyfert-type nucleus. Moreover, the [\OIII] luminosity is towards the low end of that found in Seyfert galaxies and the ratio of the 1.4 GHz radio power ($P_{\rm 1.4 GHz} = 1.1 \times 10^{23}$ W~Hz$^{-1}$) over the [\OIII] luminosity ($L_{\rm [\OIII]} = 1.3 \times 10^{39}$ erg~s$^{-1}$) is about two orders of magnitude higher than the typical ratio for Seyferts \citep{whi85,ho01}.\footnote{\citet{ho01} refer to $P_{\rm 6cm}$, but based on references from the NASA/IPAC Extra-Galactic Database (NED), $P_{\rm 1.4 GHz}$ is of the same order of magnitude as $P_{\rm 6cm}$ for B2~0722+30.} Although intrinsic reddening likely underpredicts our derived [\OIII] luminosity in the nuclear region, we do not expect this effect to be significant enough to explain the relatively high radio-power/[\OIII] ratio compared with that of Seyfert galaxies. Moreover, the emission lines in the diagnostic diagrams of \citet{vei87} were specifically chosen to lie close together in wavelength in order to minimise reddening effects on the measured line ratios. We therefore argue that intrinsic reddening effects do not alter our conclusions that B2~0722+30 has a LINER type optical AGN that is relatively weak compared to the radio source. 

From the combined radio continuum and optical emission-line properties of the AGN in B2~0722+30, we therefore classify this system as an \FRI\ radio galaxy rather than a genuine Seyfert galaxy. This implies that the low-excitation \FRI\ radio source would be hosted by the ``wrong'' kind of galaxy, given that classical radio sources are almost always hosted by early-type galaxies rather than disc-dominated galaxies.

\subsection{The ``wrong'' kind of radio galaxy}
\label{sec:wrong}

As discussed in detail in Sect. \ref{sec:introduction}, classical double-lobed radio sources are almost always hosted by early-type galaxies, generally ellipticals \citep[e.g.][]{ver01}. The occurrence of an \FRI\ radio source in the disc-dominated galaxy B2~0722+30 is therefore exceptional, though not unique. Two other cases of disc-dominated radio galaxies have been studied in detail. The extended radio source 0313-192 is a classical \FRI\ source that is hosted by a spiral galaxy \citep{led98,led01,kee06}. NGC~612 is an S0 galaxy with a large star-forming \HI\ disc, which hosts the powerful \FRI/\FRII\ radio source PKS~0131-36 \citep{ver01,emo08}. Although the linear size of B2~0722+30 is significantly smaller than that of 0313-192 and NGC~612 and its total radio power is one (0313-192) to two (NGC~612) orders of magnitude lower, the fact that all three radio sources occur in a galaxy with a prominent disc and managed to escape the optical boundaries of their hosts allows for a more detailed comparison between these systems.

The fact that the occurrence of extended radio sources in disc-dominated host galaxies is extremely rare means that these three radio galaxies are excellent systems to investigate in detail host galaxy properties or environmental effects that may be important regarding the triggering and/or evolution of radio-loud AGN. As explored by \citet{kee06} and \citet{emo08}, both 0313-192 and NGC~612 share some striking similarities that could possibly be related to the occurrence of a classical double-lobed radio source in this ``wrong'' kind of host environment: NGC~612 and 0313-192 both contain a relatively luminous bulge compared with typical spiral galaxies, which generally reflects the presence of a relatively massive central black hole \citep[e.g.][]{kor95,geb00,fer00}; in both NGC~612 and 0313-192 the radio jet axis is oriented roughly perpendicular to the disc of the galaxy (at least within $\sim 30^{\circ}$), which is likely the direction of least resistance from the ISM; in addition, both galaxies show indications that tidal encounters may have occurred, which could have resulted in enhanced fuelling of the central engine. 

\begin{figure*}
\centering
\includegraphics[width=0.92\textwidth]{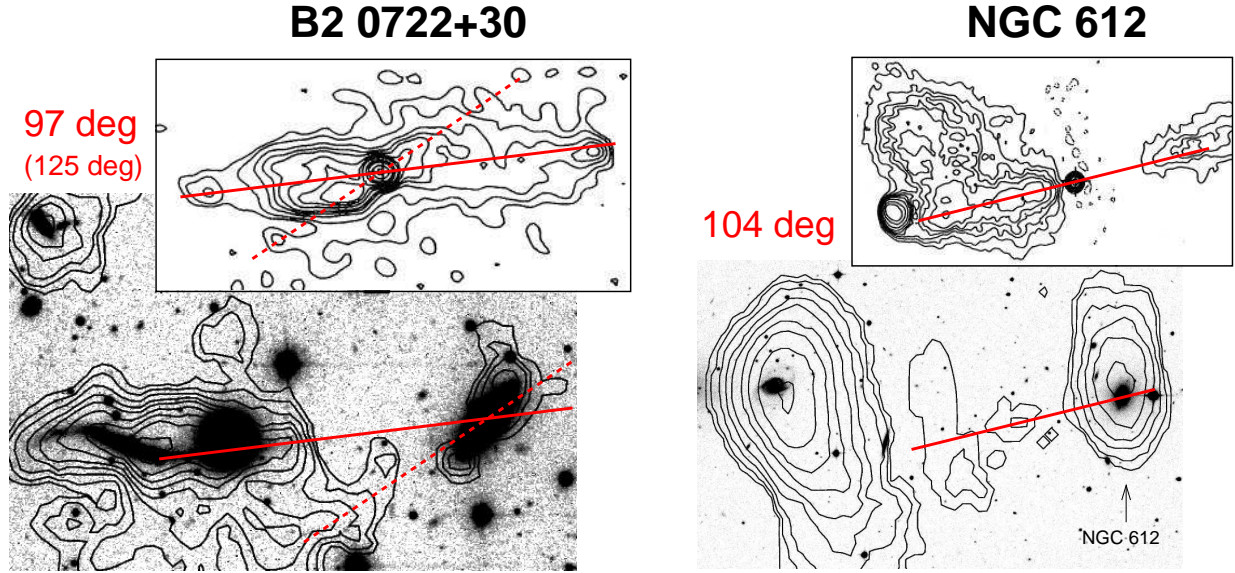}
\caption{{\sl Left:} Apparent alignment of the radio jets/lobes of B2~0722+30 in the direction of the interacting galaxy pair MRK~1201/UGC~3841a. The contours in the main panel show the total intensity \HI\ map, while those in the inset show the radio continuum \citep[taken from][]{fan86}. The solid line shows the original direction of the radio lobes (PA = 97$^{\circ}$), which is towards the extended optical tail of the distorted and elongated disc galaxy UGC~3841a. In the inner region (at kpc-scale distance from the nucleus), more recently ejected radio plasma apparently propagates roughly along PA = 124$^{\circ}$, which is in the direction of the \HI\ bridge in between B2~0722+30 and the MRK~1201/UGC~3841a pair (dashed line). {\sl Right:} a similar alignment is seen for NGC~612, where the most recently ejected radio plasma propagates in the direction of a large-scale HI bridge that extends towards a gas-rich companion galaxy \citep{emo08}. Again, the contours in the main panel show the total intensity \HI\ map, those in the inset show the radio continuum \citep[taken from][]{mor93} and the solid line indicates the position-angle of the radio lobe.}
\label{fig:jetalignment}
\end{figure*}

For B2~0722+30 there is no clear evidence for the first two possibilities. B2~0722+30 appears to be a relatively small spiral galaxy (Sect. \ref{sec:spiralhost}). The total stellar population estimated from our SED modelling in the central kpc-scale region of the bulge -- although uncertain -- indicates that the bulge is not extremely massive. Although the power of the radio source in B2~0722+30 is one to two orders of magnitude lower than that of 0313-192 and NGC~612 and therefore not necessarily abnormal in the presence of a substantially smaller bulge region and lower mass black-hole, the average-mass bulge in B2~0722+30 does not provide an obvious explanation for why this particular disc-galaxy -- compared with other spiral galaxies -- contains an extended radio source. The radio jets in B2~0722+30 are also not aligned perpendicular to the disc of the host galaxy. The large-scale radio axis lies at a position angle (PA) of about $52^{\circ}$ from the rotation axis of the disc. VLA-A resolution images of the radio continuum by \citet{fan86} show a distorted jet-lobe structure, indicating that the radio source changed direction over its lifetime \citep[see also][]{rui84}. The jet direction in the inner kpc-scale region of the radio structure -- tracing the more recently ejected jet material -- lies even closer to the disc, at an angle of $\sim 81^{\circ}$ from the disc's rotation axis. 

It is interesting to note, however, that the orientation of the large-scale radio axis is aligned in the direction of the \HI-rich MRK~1201/UGC~3841a pair. This is visualised in Fig. \ref{fig:jetalignment} {\sl (left)}. Moreover, the orientation of the inner part of the radio jets is aligned in approximately the same direction as the \HI\ bridge between B2~0722+30 and the MRK~1201/UGC~3841a pair. Although this could be merely a coincidence, we note that a similar alignment occurs in NGC~612, where the more recent radio-lobe material is aligned in the direction of an \HI\ bridge towards a companion galaxy (Fig \ref{fig:jetalignment} - {\sl right}). Perhaps a gas-rich interaction between B2~0722+30 and MRK~1201/UGC~3841a (as well as between NGC~612 and its companion) somehow cleared a path for the radio source to escape the host galaxy with minor resistance from the ISM, possibly in combination with an induced precession of the accretion-disc/jet-axis in this direction. In a future paper, we will investigate in more detail whether an alignment between the radio axis and the direction of \HI-rich companions or tidal material occurs also in other nearby radio galaxies.

\subsection{Interactions and radio source triggering}
\label{sec:triggering}

Interactions likely occurred between B2~0722+30 and nearby companions, which could be related to the triggering of the radio source in B2~0722+30 and/or to the fact that the radio source managed to escape the optical boundaries of the host galaxy. The interacting pair MRK~1201/UGC~3841a west of B2~0722+30 comprises galaxies that appear to be in ongoing interaction, based on their \HI\ properties. The faint \HI\ bridge in between B2~0722+30 and the MRK~1201/UGC~3841a system suggests that B2~0722+30 is involved in the interaction. The distribution of \HI\ gas in the MRK~1201/UGC~3841a pair in the vicinity of B2~0722+30 is very similar to that of the NGC~5058 group, where an \HI-bridge between two galaxies winds around one side of a shell elliptical \citep{guh90}. Further tentative tidal features that may indicate that B2~0722+30 has been involved in recent interactions are the \HI\ extension at the NW-tip of the disc of B2~0722+30 in the direction of a possible faint companion ("X1" in Fig. \ref{fig:Bband_B20722}) and -- if confirmed -- the fainter \HI\ tail further north. 

It is likely that galaxy interactions perturbed the gas in the central part of the disc of B2~0722+30. This could have resulted in a loss of angular momentum of the gas so that it may have been deposited on to the central black-hole and triggered the radio source \citep[see e.g.][]{lin88}. Recent studies by \citet{kuo08} and \citet{tan08} show that \HI-rich galaxy interactions are generally also associated with Seyfert galaxies, much more than with their non-active counterparts. We would like to note, however, that in an ongoing \HI\ study of a larger sample of nearby low-power radio sources (all hosted by early-type galaxies) we do not see evidence in \HI\ for ongoing gas-rich interactions associated with extended \FRI\ radio sources \citep[Emonts et al. in preparation; see also][]{emo07,emo06thesis}. In fact, recent studies suggest that the triggering mechanism for most \FRI\ radio sources -- or low-excitation radio-AGN in general -- may be related to Bondi accretion of hot circum-galactic gas on to the central black hole \citep[see e.g.][]{all06,har07,bal08}. Although our data cannot rule out a similar accretion mechanism for B2~0722+30, the results presented in this paper suggest that -- at least in certain systems and under the right conditions -- low-power \FRI\ radio-AGN can occur in systems that are undergoing gas-rich galaxy interactions. Whether or not the interactions associated with B2~0722+30 are directly responsible for the triggering of the AGN through fuelling of the black hole with low angular momentum gas needs to be investigated by studying the kinematics of the gas in the central region at higher resolution. Further investigation into this matter is important for understanding the triggering mechanism of various types of AGN.

\section{Conclusions}
\label{sec:conclusions}

B2~0722+30 is a disc galaxy that hosts a classical double-lobed radio source. Using \HI\ radio synthesis observations, deep optical imaging, stellar population synthesis modelling and emission-line diagnostics we investigated in detail the properties of the host galaxy and radio-loud AGN as well as the environment of B2~0722+30. The most important conclusions from this paper are:
\begin{itemize}
\item{B2~0722+30 contains a regular,  rotating disc of both neutral hydrogen and ionized emission-line gas. A reddened young stellar population is present throughout this disc and emission-line ratios are characteristic of \HII\ regions, which indicates that star formation is ongoing in the disc. The high FIR luminosity of B2~0722+30 may originate from dust heating by the young stellar population;}\\
\item{The AGN characteristics classify B2~0722+30 as an \FRI\ radio galaxy rather than a Seyfert galaxy; the radio source has an \FRI\ morphology, the optical AGN shows emission-line ratios characteristic of a LINER nucleus and the [\OIII] luminosity, in particular when compared to the radio power, is relatively low;}\\
\item{The presence of a classical double-lobed radio source in a disc-dominated host galaxy is extremely rare in the nearby Universe. Radio galaxies like B2~0722+30 therefore provide an excellent opportunity to identify host galaxy properties and environmental effects that could be related to the triggering mechanism of radio-loud AGN;}\\ 
\item{B2~0722+30 has an \HI-rich environment, with several gas-rich galaxies that appear to be interacting. We argue that interactions between B2~0722+30 and nearby systems could be related to the triggering of the radio source. However, more detailed kinematical studies of the gas in the nuclear region of B2~0722+30 are necessary to confirm this and to better explain the physical processes involved. The alignment of the radio axis in the direction of the gas-rich galaxy pair MRK~1201/UGC~3841a offers the interesting possibility that an interaction with this pair may be related to the fact that the radio source managed to escape the optical boundaries of the host galaxy.}
\end{itemize}

B2~0722+30 is part of a complete sample of nearby B2 radio galaxies for which we studied the large-scale \HI\ properties \citep[see][]{emo06thesis,emo07}. A detailed analysis of the complete sample will be given in a forthcoming paper.

\section*{Acknowledgments}

We would like to thank Jacqueline van Gorkom and Thijs van der Hulst for their help and useful discussions. Also thanks to Katherine Inskip for her help on the IDL-code for the SED modelling. BE acknowledges the Netherlands Organisation for Scientific Research (NWO) for funding part of this project under Rubicon grant 680.50.0508. BE also thanks Columbia University, the Kapteyn Astronomical Institute (University of Groningen), the University of Sheffield and ASTRON for their hospitality during parts of this project. This research made use of various resources: The National Radio Astronomy Observatory is a facility of the National Science Foundation operated under cooperative agreement by Associated Universities, Inc. The WHT is operated on the Island of La Palma by the Isaac Newton Group in the Spanish Observatorio del Roque de Los Muchachos of the Instituto de Astrofisica de Canarias. The Michigan-Dartmouth-MIT Observatory at Kitt Peak is owned and operated by a consortium of the University of Michigan, Dartmouth College, Ohio State University, Columbia University and Ohio University. The NASA/IPAC Extragalactic Database (NED) is operated by the Jet Propulsion Laboratory, California Institute of Technology, under contract with the National Aeronautics and Space Administration.

\bibliographystyle{mn2e} 
\bibliography{auth_total_B20722} 

\label{lastpage}

\end{document}